\documentclass[twocolumn]{webofc}

\usepackage[varg]{txfonts}   
\usepackage{bm}
\usepackage{amsmath}
\usepackage{physics}
\usepackage{algorithm2e}
\begin{document}
\title{Quantum Computing for Multi Period Asset Allocation}
%
%

\author{
    \firstname{Queenie} \lastname{Sun}\inst{1,2}\fnsep\thanks{\email{sunq6@rpi.edu}}
    \and
    \firstname{Nicholas} \lastname{Grablevsky}\inst{1,2}\fnsep\thanks{\email{grabln@rpi.edu}}
    \and
    \firstname{Huaizhang} \lastname{Deng}\inst{3}
    \and
    \firstname{Pooya} \lastname{Azadi}\inst{3}
}
\institute{
Department of Computer Science, Rensselaer Polytechnic Institute, 110 8th St., Troy, NY, USA
\and
Rubiqon Inc
\and
The Rock Creek Group, 1133 Connecticut Avenue NW Washington, DC 20036
}

\abstract{%
    Portfolio construction has been a long-standing topic of research in finance. The computational complexity and the time taken both increase rapidly with the number of investments in the portfolio. It becomes difficult, even impossible for classic computers to solve. Quantum computing is a new way of computing which takes advantage of quantum superposition and entanglement. It changes how such problems are approached and is not constrained by some of the classic computational complexity. Studies have shown that quantum computing can offer significant advantages over classical computing in many fields. The application of quantum computing has been constrained by the unavailability of actual quantum computers. In the past decade, there has been the rapid development of the large-scale quantum computer. However, software development for quantum computing is slow in many fields. In our study, we apply quantum computing to a multi-asset portfolio simulation. The simulation is based on historic data, covariance, and expected returns, all calculated using quantum computing. Although technically a solvable problem for classical computing, we believe the software development is important to the future application of quantum computing in finance. We conducted this study through simulation of a quantum computer and the use of Rensselaer Polytechnic Institute’s IBM quantum computer.
    
}
\maketitle
\section{Introduction}\label{intro}
The crux of quantum computing lies in leveraging two concepts from quantum mechanics: superposition and entanglement. A classical computer is composed of bits, which can be in a state of either a 0 or a 1. Quantum computers are composed of quantum bits, or qubits, which can be in a state of superposition. Superposition is the notion that the state of a qubit is in a state of \textit{both} 0 \textit{and} 1. When qubits are measured, their quantum state collapses to either 0 or 1, providing the output of the computation. Entanglement is the principle that multiple qubits, despite any form of separation, are correlated. A change in one inherently provokes a change in the other \cite{prashant2007superposition}\cite{gaasbeek2010introductory}. Through these quantum mechanics principles, quantum algorithms see their advantage over a wide variety of problems. 

The advantages of quantum computing were theorized long before any physical quantum computers were created. Numerous studies have explored this subject, one of the most famous being Shor’s algorithm, which was developed in 1994 and has been proven to factor numbers exponentially faster than classical computers \cite{Shor_1997}. In the past decade, hardware technology has advanced rapidly, enabling the development of actual quantum computers. Although still limited by hardware constraints, these computers have allowed for the testing of this algorithm, where it has been proven to work in practice \cite {Willsch_2023}. Quantum computing is being transitioned into quantum reality, enabling the validation or refutation of theoretical algorithms. Today is a remarkably unique era of quantum exploration that will lead to noteworthy progress in both hardware and software, impacting all industries.  

Quantum computers continue to highlight their efficacy through their ability to evaluate potential solutions simultaneously \cite{abbas2023quantum}. This gives them a distinct advantage in solving problems that involve optimization or large-scale simulations, which is particularly useful in finance. However, the software in quantum application in finance is extremely limited. In this study, we explore the application of quantum computing in solving a multi-period asset allocation problem. Although the simple case in this study can be solved easily by classical computers, our goal is to contribute to the software foundation for the future of quantum applications in finance. 

\section{IBM Quantum System One}\label{sec:cancellation}

In this study, we utilized IBM's Quantum System One, a state-of-the-art quantum computer housed at Rensselaer Polytechnic Institute. As students at Rensselaer, we have full and prioritized access to this on-campus quantum computer, which significantly reduces job queue wait times, enabling faster code deployment and testing compared to studies relying on public quantum computers.

The quantum computer employed in this study features 127 qubits, powered by an IBM Quantum 'Eagle' Processor, which has demonstrated the ability to produce accurate calculations beyond brute-force simulation methods. The quantum computer functions by changing positions of qubits in the quantum computer. It is stored in a refrigerator that is nearing 0 Kelvin, theoretically the lowest temperature. The quantum computer is linked to the IBM cloud, facilitating seamless access and management of quantum jobs from anywhere on campus. Before sending out jobs to the physical quantum computer, we conducted extensive testing on IBM's quantum simulator. This allowed us to refine our code and algorithms, ensuring that we are satisfied with the simulation results before sending it to the actual quantum computer. 

\begin{figure}[h]
    \centering
    \includegraphics[width=0.5\textwidth]{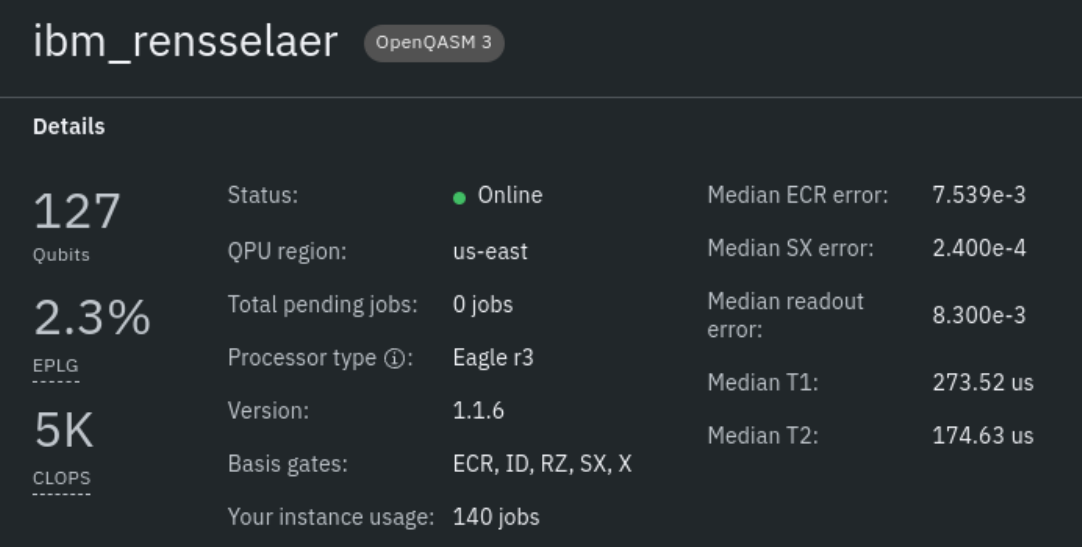}
     \caption{For a visual understanding, a job gets sent to this quantum computer on the IBM cloud, which runs the circuit that it is sent.}
    \label{fig:generated_asset}
\end{figure}
\begin{figure}[h]
    \centering
    \includegraphics[width=0.5\textwidth]{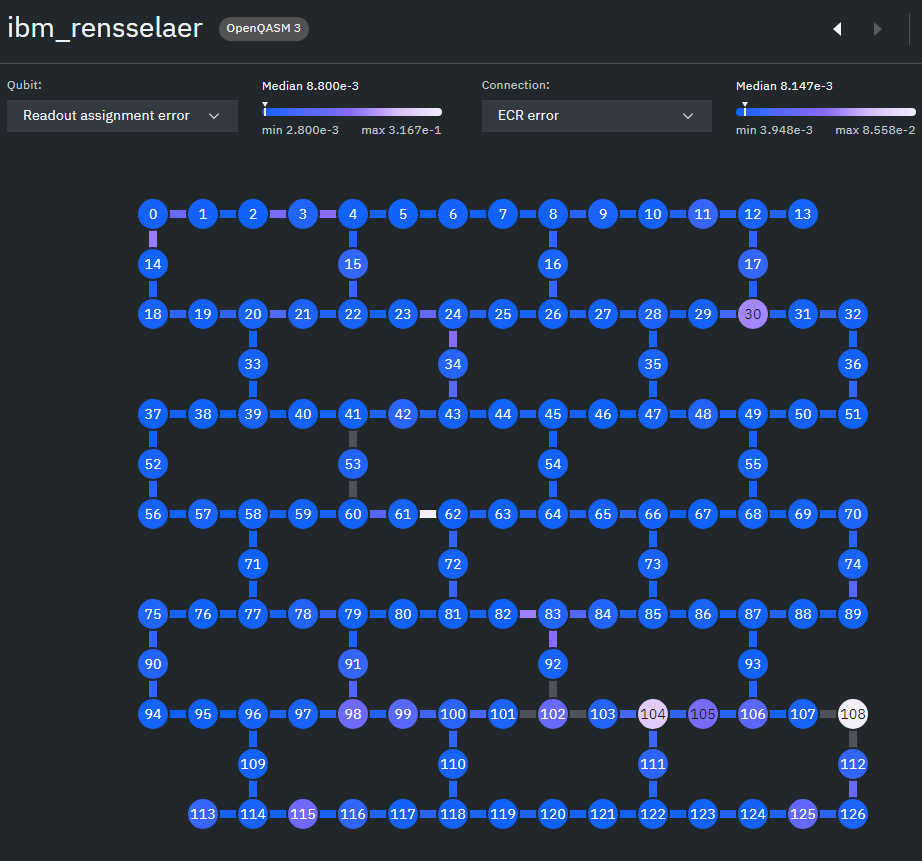}
     \caption{Depicted above is a visualization of the qubits in ibm\_rensselaer.}
    \label{fig:generated_asset}
\end{figure}

\section{Multi-Asset Multi-Period Asset Allocation }\label{sec:cancellation}

In this study, the portfolio consisted of three asset classes, US equities, international equities, and global fixed income. We used the past 20 years’ historical data of S\&P 500, MSCI All Country World Index ex US, and Bloomberg Global Aggregate indices as the proxy to calculate the volatilities and covariance matrix. The expected returns are chosen to be 10\%, 10\%, and 6\% respectively for the three asset classes. The simulations assume multivariate log-normal distribution and are done monthly. 

The classical method for performance simulation involves calculating the multivariate normal distribution. In contrast, the quantum computer executes a specialized quantum circuit that produces a normal distribution discretized over a specified number of qubits. In our study, we allocate 3 qubits to each of the three asset classes, resulting in a qubit array of [3, 3, 3]. For more complex scenarios or when higher accuracy is required, additional qubits can be utilized. The distribution bounds are set to three standard deviations above and below the expected returns.

With the qubit array, expected return $\mu$, the covariance matrix $\Sigma$, and bounds defined, the quantum circuit for the multivariate normal distribution is generated. A new quantum circuit is created with a size equal to the sum of qubits in the array. Building a quantum circuit involves first defining these qubits, which serve as quantum analogs of classical bits, but with the unique ability to exist in superpositions of states. The circuit is then constructed by applying a sequence of quantum gates—analogous to classical logic gates—which manipulate the qubits' states through operations like superposition, entanglement, and interference. These operations are carefully orchestrated to perform the quantum algorithm, ultimately leading to the desired computational outcome when the qubits are measured. These two quantum circuits are then combined. After measuring this combined circuit, a quantum circuit ready for execution is obtained.

Each execution runs 120 shots of the quantum circuit, with each shot representing the returns for one month. Therefore, each execution simulates 10 years of performance. After execution, we convert the qubit measurements into returns and compare the covariance matrix of the input with that of the output to ensure that the quantum computer's inherent errors are minimal. A sample of the difference is provided below.

\[
\mathbf{\Delta\Sigma} = \begin{pmatrix}
-0.37 & -0.05 & 1.53 \\
-0.05 & 0.01 & 1.61 \\
1.53 & 1.61 & 0.52
\end{pmatrix} \times 10^{-4}
\]
\begin{figure}[h]
    \centering
    \includegraphics[width=0.5\textwidth]{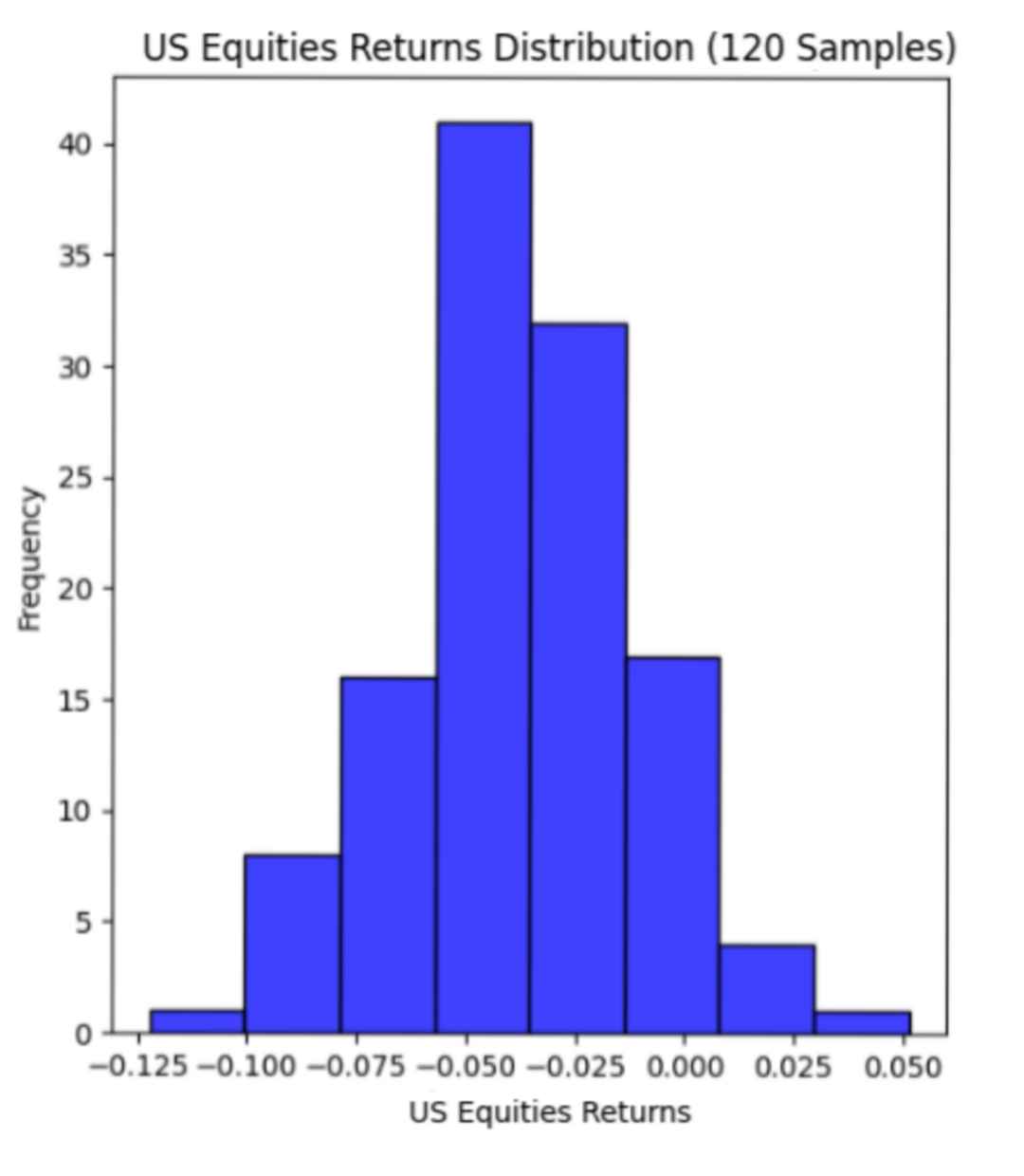}
     \caption{The distribution of US equity returns in one execution.}
    \label{fig:generated_asset}
\end{figure}

With the generated return data, we implemented different types of rebalancing rules including monthly, quarterly, semi-annually, and annually.

\section{Quantum Reality and Conclusion}

In our examples, we use 3 qubits per asset class. While theoretically, using more qubits would enhance simulation accuracy, we tested this approach with additional qubits. The main limitation of this method lies in the time required to build the quantum circuit.

On the quantum emulator, running a qubit array of [3, 3, 3] allows the circuit to be built within a few seconds. However, increasing the array to [4, 4, 4] results in a circuit that takes approximately one minute to construct. The most extensive test we performed used a qubit array of [5, 5, 5], which required about 20 minutes to build.

When running the circuit on a physical quantum computer, the [3, 3, 3] qubit array completed in just a few seconds. With the [4, 4, 4] qubit array, the build time increased to 30 seconds, and with the [5, 5, 5] qubit array, the quantum computer failed after around 6 minutes.

In addition to this limitation, the current state of quantum computing software is still quite primitive. Functional circuits that work well on an emulator often exhibit significant discrepancies when run on a physical quantum computer.

Over the past decade, quantum computers have evolved from a few qubits to thousands. Although quantum computers are not yet fully practical for widespread use, they have become reasonably accessible, enabling students to conduct research in various fields, including finance in our case. As quantum hardware development accelerates in the coming decade, software development will need to keep pace. We believe that the availability of quantum computers to college students is crucial for establishing a quantum software community. This community will develop applications that fully harness the potential of quantum computers and help bring quantum supremacy closer to reality. With the on-campus quantum computer and STEM focused students, Rensselaer Polytechnic Institute is uniquely positioned to develop such community in this crucial era of quantum computing.

\section*{Acknowledgements}
We would like to thank the Rensselaer Polytechnic Institute (RPI) Division of the Chief Information Officer (DotCIO) for providing access to the quantum computer which has been recently acquired with a generous donation from Curtis R. Priem.

%

\bibliography{main}

\end{document}